\documentclass[aps,epjc,twocolumn,floatfix,superscriptaddress,nofootinbib,showpacs,showkeys,preprintnumbers]{revtex4}
\usepackage{graphicx}
\usepackage{dcolumn}  
\usepackage{bm}       
\usepackage{palatino}

\preprint{preprint number}
\usepackage{epsfig}
\begin{document}
\title{ Simple parameterization of nuclear attenuation data} 
\author{N.~Akopov}
\author{L.~Grigoryan}
\author{Z.~Akopov \\
Yerevan Physics Institute, Br.Alikhanian 2, 375036 Yerevan, Armenia }
\begin {abstract}
Based on the nuclear attenuation data obtained by
the HERMES experiment on nitrogen and krypton nuclei,
it is shown that the nuclear attenuation $R_M^{h}$ can be 
parametrised in a form of a linear polynomial $P_1=a_{11}$ + $\tau a_{12}$,
where $\tau$ is the formation time, which depends on the energy of the virtual 
photon $\nu$ and fraction of that energy $z$ carried by the final hadron. Three
widely known parameterizations for $\tau$ were used for the performed fit.
The fit parameters $a_{11}$ and $a_{12}$ do not depend on $\nu$ and $z$.
\end {abstract}
\maketitle
\section{Introduction}
\normalsize
Semi-inclusive Deep Inelastic Scattering (DIS) of leptons on 
nuclear targets is widely used for studies of the hadronization
process~\cite{osborn,A2,A7,A8}. It is most effective to observe at moderate 
energies of the virtual photon, when the formation time of the hadron is 
comparable with the nuclear radius.\\
Hadronization is the process through which partons, created in an
elementary interaction, turn into the hadrons that are observed experimentally.
According to theoretical estimates the hadronization process occurs over length
scales that vary from less than a femtometer to several tens of femtometers.
Hadronization in a nuclear environment is interesting due to two main reasons.
First of all, it allows to study the parameters governing this process on the 
early stage; on the other hand, it can provide initial conditions
for the investigation of hadronization in hot nuclear matter, which
arises in high energy ion-ion collisions.\\ 
The most convenient observable measured experimentally for this process is the 
Nuclear Attenuation (NA) ratio, which is a ratio of multiplicities on nucleus
and deuterium (per nucleon). Next step is to find
a variable which allows to present the NA ratio in the most simple form.\\
We propose the formation time $\tau$ as the best variable
for the NA and show that the data can be
parametrised in the form of a linear polynomial 
$P_1=a_{11} + \tau a_{12}$. 
Formation time $\tau$ depends on the energy of photon
$\nu$ and fraction of this energy $z=E_h/\nu$ carried by the final hadron with
energy $E_h$. Three
widely known parameterizations for $\tau$ were used for the fit procedure.
The parameters $a_{11}$ and $a_{12}$, obtained from the fit, do not depend on $\nu$ and $z$.
They are functions of the prehadron-nucleon and hadron-nucleon
cross sections and the atomic mass number.\\
The NA data for pions on nitrogen and for identified hadrons on krypton nuclei obtained by 
the HERMES experiment~\cite{A7,A8} were used to perform the fit.\\
This paper is organized as follows. Nuclear attenuation in absorption
model is presented in the next section. In Section 3 it is discussed
a choice of the appropriate form for the variable $\tau$. Section 4 presents results of
the fit. Conclusions are given in Section 5.
\section{Nuclear attenuation in absorption model}
\normalsize
The semi-inclusive DIS of lepton on nucleus of atomic mass number 
A is:
\begin{eqnarray}
          l_i + A \rightarrow l_f + h + X,
\end{eqnarray}
where $l_i (l_f)$ are the initial (final) leptons, and h is the hadron 
observed in the final state. This process is usually investigated in
terms of the NA ratio, which is frequently defined as a function of two variables, 
$\nu$ and $z$
\footnote{In fact, $R_M^{h}$ also depends on the 
photon virtuality $Q^2$ and on the square
of the  hadron transverse momentum $p_t^{2}$, in respect to the virtual photon direction. 
However, from the experimental data, it is known that $R_M^{h}$
is a much sensitive function of $\nu$ and $z$ in comparison with $Q^2$ and $p_t^{2}$.}
:
\begin{eqnarray}   
       {R_M^{h}(\nu,z) \approx 2d\sigma_A(\nu,z)/Ad\sigma_D(\nu,z)}.
\end{eqnarray}
Usually it is  investigated  at precise values of one variable and average
values of another.\\
In case where the $\nu$ - dependence is studied,  
\begin{eqnarray}
       {R_M^{h}(\nu,\langle z \rangle) = 2d\sigma_A(\nu,\langle z \rangle)/Ad\sigma_D(\nu,\langle z \rangle)},
\end{eqnarray}
where $\langle z \rangle$ are the average values of $z$ for each $\nu$ bin.
And for $z$ - dependence
\begin{eqnarray}
       {R_M^{h}(\langle \nu \rangle ,z) = 2d\sigma_A(\langle \nu \rangle,z)/Ad\sigma_D(\langle \nu \rangle,z)},
\end{eqnarray}
where $\langle \nu \rangle$ are the average values of $\nu$ for each $z$ bin.

In this work we adopt a model, according to which the origin
of NA is the absorption of the prehadron (string,
dipole) and final hadron in the nuclear medium. In that case 
NA ratio has the following form:
\begin{eqnarray}
{R_M^{h}=\int{d^2b} \int_{-\infty}^{\infty}
{\rho(b,x)[W(b,x)]^{(A-1)}dx}},
\end{eqnarray}
where $W(b,x)$
is the probability that neither the prehadron produced at the DIS point
$(b,x)$ (where $b$ is impact parameter and $x$ the longitudinal coordinate), 
nor the hadron produced at point $(b,x')$ is absorbed by one nucleon in
the nucleus; $\rho$ is the nuclear density function with a normalization
condition:
\begin{eqnarray}
\nonumber
{\int{\rho(r)d^{3}r}=1}.
\end{eqnarray}
For $W(b,x)$ we use the one scale model proposed in Ref.~\cite{A1}:
\begin{eqnarray}
{W(b,x)=1-\sigma_q \int_{x}^{\infty}{P_q(x'-x)\rho(b,x')dx'}-} \\
\nonumber
{\sigma_h \int_{x}^{\infty}{P_h(x'-x)\rho(b,x')dx'}},
\end{eqnarray}
where $\sigma_q$ and $\sigma_h$ are the inelastic cross sections for
prehadron-nucleon and hadron-nucleon interactions, respectively.
$P_q(x'-x)$ is the probability that on distance $x'-x$ from DIS point
particle is a prehadron and
$P_h(x'-x)$ is the probability that particle is a hadron.
\begin{eqnarray}
{P_h(x'-x)=1-P_q(x'-x)}.
\end{eqnarray}
In analogy with survival probability for a particle having
lifetime $\tau$ in a system where it travels a distance $x'-x$
before decaying, $P_q(x'-x)$ can be expressed in form:
\begin{eqnarray}
{P_q(x'-x)=\exp[-(x'-x)/\tau]},
\end{eqnarray}
where $\tau$ is the formation time.
Substituting expressions for $P_q(x'-x)$ and $P_h(x'-x)$ in eq.(6)
we obtain: 

\begin{eqnarray}
{W(b,x)\approx{1-\sigma_h \int_{x}^{\infty}{\rho(b,x')dx'}
+{\tau(\sigma_h-\sigma_q)\rho(b,x)}}}\\
\nonumber
{\approx{w_1(b,x)+\tau(\nu,z)w_2(b,x)}}.
\end{eqnarray}  
$W$ depends on $\nu$ and $z$ only by means of
$\tau(\nu,z)$.\\
In more detail, the formation time in string models can be divided in two parts
(see, for instance, two scale model presented in Ref~\cite{A2,A3}). 
First past is the constituent formation time $\tau_c$,
which defines the time elapsed from the moment of the DIS untill the production of the 
first constituent of the final hadron. Second time interval begins
with the production of the first constituent until the second one, which coincides with 
the yo-yo\footnote{The yo-yo 
formation means that a colorless system with valence contents and
quantum numbers of the final hadron is formed, but without its 
"sea" partons.}
or final hadron production. 
Comparison with the experimental data
shows that in the second interval, the prehadron-nucleon cross section has
values close to the hadron-nucleon cross section $\sigma_h$. If the 
difference between these cross-sections is neglected, the model 
is reduced to one scale model with $\tau = \tau_c$.
In case of the improved two scale model Ref~\cite{A3}, the prehadron-nucleon cross section 
reaches hadron-nucleon cross section value during a time interval 
$\tau = \tau_c + c\Delta\tau$, where $\Delta\tau = z\nu/\kappa$,
$\kappa$ is the string tension,
c is the free parameter which defines from fit. In Ref~\cite{A3} it is
shown that $c \ll 1$. 
Transition to the one scale model takes place at c=0 and corresponds to
$\tau = \tau_c$.
One should note that any complicated absorption string model, in some approximation, can be 
reduced to the one scale model presented in eq.(6) and (9). 
Substituting $W(b,x)$ in $R_M^{h}$ we obtain:
\begin{eqnarray}
\nonumber
{R_M^{h} \approx{\int{d^2b} \int_{-\infty}^{\infty}{\rho(b,x)
(w_1+\tau w_2)^{(A-1)}dx}}}\\
\approx{a_{i1} + \tau a_{i2} + \tau^2 a_{i3} + {}\cdots},
\end{eqnarray}
where $i$ is the maximal power of $\tau$ with which we are limited.
The coefficients $a_{ij}$ depend on $A, \sigma_q, \sigma_h$ and nuclear
density. This means, that $a_{ij}$ vary for different
nuclei. For each nucleus $a_{ij}$ are the same for hadrons with equal
values of $\sigma_q$ and $\sigma_h$ (for instance for pions\footnote{
we do not mention the electric charge of pions, because
cross sections of differently charged pions with nucleons,
which are of interest to us, are equal.} 
and negatively charged kaons).
Although $R_M^{h}$ is a polynomial of $\tau$
with maximal power $A-1$, it is expected that
$a_{i1} > a_{i2} > a_{i3} > {}\cdots$.
For fitting of the NA data we use three 
expressions for $R_M^{h}$ as first, second and third order
polynomials of $\tau$ :
\begin{eqnarray}
{P_1={a_{11} + \tau a_{12} }},
\end{eqnarray}
\begin{eqnarray}
{P_2={a_{21} + \tau a_{22} + \tau^2 a_{23} }},
\end{eqnarray}
\begin{eqnarray}
{P_3={a_{31} + \tau a_{32} + \tau^2 a_{33} + \tau^3 a_{34}}}.
\end{eqnarray}
In order to get the information on the influence of highest order of
polynomials, also $P_4$ polynomial was checked (see section 4).
\section{Formation time}
\normalsize
Equation (10) shows, that within our approximation, $R_M^{h}$ is a 
function of $\tau$ only. In
this section we will discuss the physical meaning and possible
expression of the formation time $\tau$.
There are different definitions for the formation time. 
We define formation time as a time scale within which
the prehadron-nucleon cross section reaches the value of the
hadron-nucleon one.
In the literature there are three qualitatively different
definitions for $\tau$. In the first extreme case it is assumed
that $\tau = 0$ (Glauber approach). In the second extreme case
$\tau \gg r_A$, where $r_A$ is the nuclear radius (energy loss 
model~\cite{energyloss}). And at last, in our opinion more realistic definition 
of the formation time, as a function
of $\nu$ and $z$ which can change from zero up to values
larger than $r_A$. Experimental data seem to confirm that
for moderate values of $\nu$ on the order of
$10 GeV$ the formation time
is smaller than the nuclear size, i.e. the hadronization takes
place within the nucleus. We will use the formation times according to
the third definition mentioned above.
Following expressions are used:\\
1. Formation time for the leading hadron~\cite{A4}, which 
follows from the energy-momentum conservation law 
\begin{eqnarray}
{\tau_{lead}=(1-z)\nu/\kappa} ,
\end{eqnarray}
where $\kappa$ is the string tension (string constant) with numerical 
value $\kappa = 1 GeV/fm$. Indeed, eq. (14) presents formation time
for the hadron produced on the fast end of the string or, which is the same, 
for the last hadron produced from string. 
Hadrons can be produced along whole length of the
string. Among them, the hadrons produced on the fast end have
better chance to avoid interactions in the nucleus.\\
2. Formation time for the fast hadron, which is composed
of characteristic formation
time of the hadron $h$ in its rest frame $\tau_0$ and
Lorentz factor (see Ref.~\cite{A1})
\begin{eqnarray}
{\tau_{Lor}=\tau_0\frac{E_h}{m_h}=\tau_0\frac{z\nu}{m_h}} ,
\end{eqnarray}
where $E_h$ and $m_h$ are the energy and mass of the hadron $h$,
respectively. Let us briefly discuss the factor $\tau_0$.
Some authors assume (see Ref.~\cite{A5}) that $\tau_0$
is a universal quantity which does not depends on 
the hadron type. If this assumtion is correct, then
$\tau_{Lor}$ for the kaons is approximately 3.5 times shorter
than for the pions at the same values of $\nu$ and $z$, although
the experiment gives $R_M^{\pi} \approx R_M^{K^{-}}$, and it is known
that $\sigma_{\pi} \approx \sigma_{K^{-}}$. Such definition
seems merely strange in framework of the absorption model. 
It appears more realistic that $\tau_0$ is proportional to $m_h$.
The reason for this is that in the string model the meson (baryon) 
is represented as a system consisting of
a quark-antiquark (diquark) and a gluonic string between them.
The energy of the system is transfered by the gluonic string from one parton to
another and back.
One full cycle lasts a period of $m_{h}/\kappa$ which we adopt
as $\tau_0$. Then $\tau_{Lor}$ is a universal quantity which does not
depend on the hadron type.\\
3. Formation time following from the Lund string model Ref.~\cite{A6}
\footnote{Note that this approximation is used only for the sake of
easy reading. For numerical calculations we use the precise expression
for $\tau_{Lund}$ following from equation $\tau_{Lund}=\tau_y-z\nu/\kappa$
with $\tau_y$ taken from eq.(4.21) of Ref.~\cite{A6}.}
\begin{eqnarray}
{\tau_{Lund}=\Bigg[{\frac{\ln(1/z^2)-1+z^2}{1-z^2}}
\Bigg]{\frac{z\nu}{\kappa}}} .
\end{eqnarray}
One should note that all three types of formation time have similar behaviour
with $\nu$, but different behaviour with $z$. At the values of $z$
typical for the HERMES kinematics the behavior of $\tau$ defined as (14) and (16)
is similar, i.e. they are decreasing with the increase of $z$, while $\tau$
defined as (15) is increasing with the increase of $z$.
\section{Results}
\normalsize
A combined fit was performed for each nucleus and all hadrons
having equal cross sections and, consequently, identically absorbed 
by nuclear matter (the same applies to their prehadrons).
It is worth reminding, that the inelastic cross sections of corresponding
hadrons with nucleons in the moderate energy range ($E_h \sim 10 GeV$)
have the following values:
$\sigma_{\pi} = \sigma_{K^{-}} = 20mb$, $\sigma_{K^{+}}
= 14mb$, $\sigma_{p} = 32mb$ and $\sigma_{\bar{p}} = 42mb$.
Combined fit was performed only for the hadrons which have equal cross-sections.
\begin{figure}[!ht]
\begin{center}
\epsfxsize=8.cm
\epsfbox{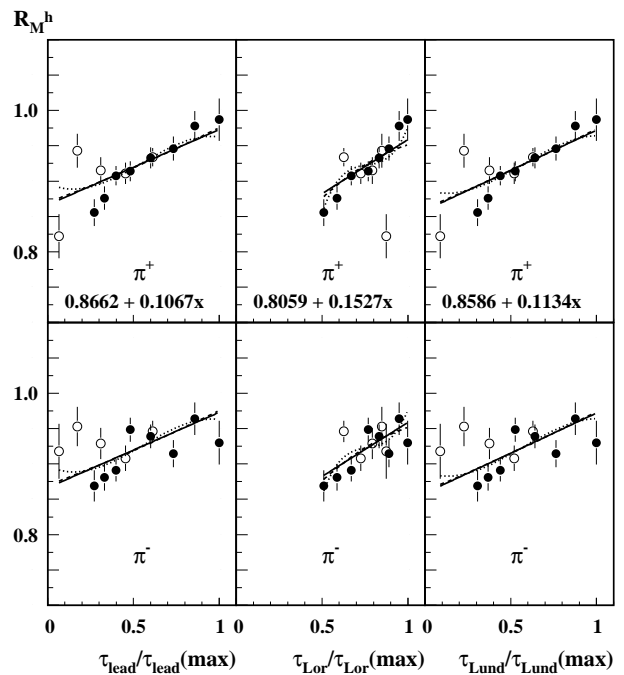}
\end{center}   
\caption{\label{xx1}
{\it The values $R_M^{h}$ on nitrogen as a function of
$\tau_{lead}$ (left two panels),
$\tau_{Lor.}$ (central two panels),
$\tau_{Lund}$ (right two panels).
Normalized values $\tau/\tau(max)$
for all $\tau$ are  used.
On upper panels $\pi^+$, on 
lower $\pi^-$ mesons are presented respectively.
Solid, dashed and dotted curves are results of 
linear, quadratic and cubic polynomial fits.
The numerical results for linear fit are noted on upper panels.
}}
\end{figure}
\begin{figure}[!ht]
\begin{center}
\epsfxsize=8.cm
\epsfbox{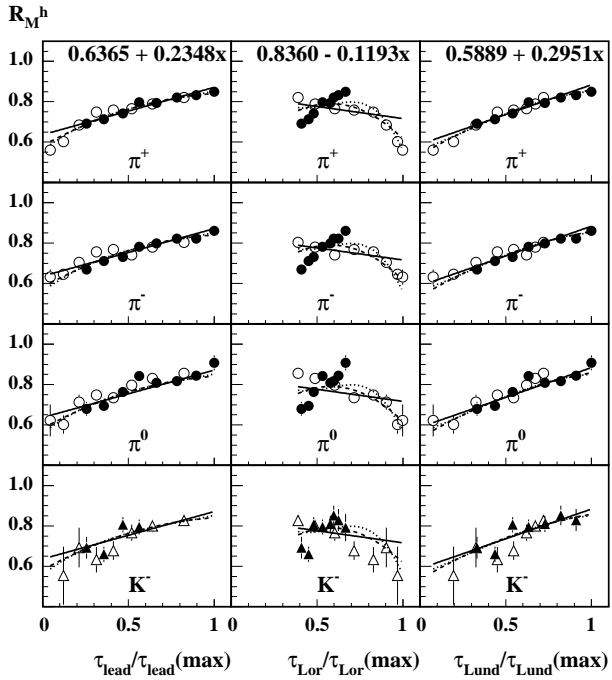}
\end{center}   
\caption{\label{xx2}
{\it The values $R_M^{h}$ on krypton as a function of
$\tau_{lead}$ (left four panels),
$\tau_{Lor.}$ (central four panels),
$\tau_{Lund}$ (right four panels).
On panels from upper to lower $\pi^+$,
$\pi^-$, $\pi^0$ and $K^-$ mesons, are presented respectively.
}}
\end{figure}
\begin{figure}[!ht]
\begin{center}
\epsfxsize=8.cm
\epsfbox{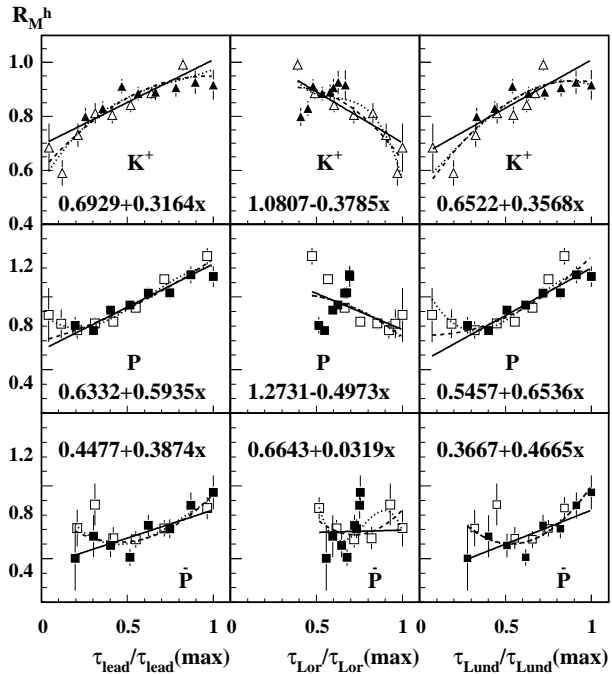}
\end{center}
\caption{\label{xx3}
{\it The values $R_M^{h}$ on krypton as a function of 
$\tau_{lead}$ (left three panels),
$\tau_{Lor.}$ (central three panels),
$\tau_{Lund}$ (right three panels).
The results for $K^+$ mesons are presented on upper panels, protons - on
middle, and antiprotons - on lower.
}}
\end{figure}
As a result, two combined fits were performed for: 
\begin{enumerate}

\item  positive and
negative pions on nitrogen (26 experimental points from~\cite{A7});

\item positive, negative, neutral pions and 
negative kaons on krypton (63 experimental points from~\cite{A8});
\item and separate fits for:
\begin{itemize}
\item positive kaons on krypton (16 experimental points~\cite{A8});

\item protons on krypton (16 experimental points~\cite{A8}) and

\item antiprotons on krypton (14 experimental points~\cite{A8}).
\end{itemize}
\end{enumerate}
The NA ratios were taken in polynomial form (11)-(13), and formation
times (lengths) as in (14)-(16). The results for the reduced $\chi^2$
denoted as $\chi^2/{d.o.f.}$ are presented in Table~\ref{tab:Table1}.
One can see that for each choice of formation time
and for each nucleus, the values of $\chi^2/{d.o.f.}$ are close 
for the polynomials $P_1$, $P_2$ and $P_3$,
which means that including higher power polynomials
of $\tau$ will not essentially improve the description of data. 
In order to test this, we calculated also $P_4$ polynomials and
obtained the values of $\chi^2/{d.o.f.}$ close to the ones in case of $P_3$.
From Table~\ref{tab:Table1} one can see that the fit gives 
unexpectedly good values for $\chi^2/{d.o.f.}$ close to unity for $\tau_{lead}$ and
$\tau_{Lund}$, and for $\tau_{Lor}$ the agreement is much worse.
Experimental points as a function of $\tau$ and results of the fit are presented in 
Fig.~\ref{xx1} for nitrogen and in Figures~\ref{xx2},~\ref{xx3} for krypton. 
Solid points correspond to the NA ratio obtained from the experimental data for 
$\nu$-dependence, open points for $z$-dependence. For convenience we have renormalized
$\tau$ to $\tau/\tau(max)$, where $\tau(max)$ are the maximum
values of $\tau$ for each set of data and each choice of $\tau$ expression.
On each of the figures the linear polynomial is presented
$a_{11} + \tau a_{12}^{'}$ with values $a_{11}$ and 
$a_{12}^{'}$ = $a_{12}\tau(max)$ corresponding to the best fit.
Solid, dashed and dotted curves correspond to $P_1$, $P_2$ and $P_3$
polynomial fit.
\section{Conclusions.}       
\normalsize
\begin{itemize}
\item Based on the performed studies one can assume that NA ratio is a function of formation
time $\tau$ only.
\item Three expressions for NA ratio as a first, second and third order 
polynomials of $\tau$ were used for these studies, and the results do not show remarkable
difference in their ability to describe the HERMES data.
\item The performed analysis with three different
expressions for the formation time $\tau$ shows that based on the obtained
values of the reduced $\chi^2$ (see Table 1), as well as on visual shape of
the curves in Figures~\ref{xx1},~\ref{xx2} and~\ref{xx3}, one can conclude that 
the two of them - $\tau_{lead}$ and $\tau_{Lund}$ are quite appropriate, and the
expression for the formation time in form of $\tau_{Lor}$ is ruled out.
\item The performed combined fits are based on the HERMES NA data using:
$\pi^+$, $\pi^-$ for nitrogen;
$\pi^+$, $\pi^-$, $\pi^0$, $K^-$ for krypton;
separate fits for $K^+$, protons and antiprotons for krypton.
\item Because of the fact that the linear function gives about the same values 
  $\chi^2/{d.o.f.}$ values as second and third order one, we assume that in our
approach the parameterization by linear function is well enough.
\end{itemize}

\vspace {-1cm}
\begin{table*}[htb]
\caption{ $\chi^2/{d.o.f.}$ from polynomial fit. Details see in text}
\label{tab:Table1}
\begin{center}
\begin{tabular}{|c|c|c|c|c|c|c|c|c|c|c|c|} \hline
\multicolumn{12}{|c}{\hskip 3.45cm \vline \hskip 0.8cm
$\tau_{lead.}$ \hskip 0.75cm \vline \hskip 0.85cm $\tau_{Lor}$
 \hskip 0.8cm \vline \hskip 0.65cm $\tau_{Lund}$} \hskip 0.7cm \vline\\
\hline
{ A}&{ $N_{exp}$}&{ $Had.$}&
{ $P_1$}&{
$P_2$}&{ $P_3$}&{ $P_1$}&{ $P_2$}&{ $P_3$}&
{ $P_1$}&{ $P_2$}&{ $P_3$}\\
\hline
{ $^{14}N     $}&{ 26}&{ $\pi^{+,-}$}&{
1.65}&{ 1.72}&{ 1.73}&{ 1.89}&{ 1.95}&
{ 1.81}&{ 1.60}&{ 1.67}&{ 1.71}\\
\hline
{ $^{84}Kr$}&{63}&{ $\pi^{+,-,0},K^-$}&{
1.67}&{ 1.32}&{ 1.31}&{ 8.71}&{ 6.62}&
{ 5.98}&{ 1.41}&{ 1.23}&{ 1.23}\\
\hline
{\ $^{84}Kr$}&{ 16}&{ $K^+$}&{
1.95}&{ 1.62}&{ 1.70}&{ 3.47}&{ 3.30}&
{ 3.14}&{ 2.78}&{ 2.34}&{ 2.52}\\
\hline
{ $^{84}Kr$}&{ 16}&{ $proton$}&{
1.25}&{ 1.24}&{ 1.04}&{ 8.50}&{ 9.08}&
{ 9.82}&{ 2.48}&{ 2.28}&{ 1.90}\\
\hline
{ $^{84}Kr$}&{ 14}&{ $antipr.$}&{
1.33}&{ 0.89}&{ 0.92}&{ 2.43}&{ 2.37}&
{ 2.04}&{ 1.50}&{ 0.94}&{ 1.03}\\
\hline
\end{tabular}
\end{center}
\end{table*}




\begin{thebibliography}{99}
\bibitem{osborn} L.S.~Osborn et al., Phys. Lett. {\bf 40B}, (1978), 1624 
\bibitem{A2}  J.Ashman et al., Z.Phys. {\bf C52}(1991) 1
\bibitem{A7}  A.Airapetian et al., Eur.Phys. J. {\bf C20} (2001) 479
\bibitem{A8}  A.Airapetian et al., Phys.Lett. {\bf B577} (2003) 37
\bibitem{A1}   A.Bialas, Acta Phys.Pol. {\bf B11} (1980) 475
\bibitem{A3}  N.Akopov, L.Grigoryan, Z.Akopov, Eur.Phys.J. {\bf C44}
(2005) 219; hep-ph/0409359 (2004)
\bibitem{energyloss} X.-N.~Wang, X.~Guo, Nucl. Phys., {\bf A696} (2001), 788;
E.~Wang, X.-N.~Wang, Phys. Rev. Lett. {\bf 89} (2002), 162301 
\bibitem{A4}  B.Kopeliovich, Phys.Lett. {\bf B243} (1990) 141
\bibitem{A5}  T.Falter et al., Phys.Lett. {\bf B594} (2004) 61
\bibitem{A6}  A.Bialas, M.Gyulassy, Nucl.Phys. {\bf B291} (1987) 793
\end{thebibliography}
\end{document}